\documentclass[12pt]{iopart}

\usepackage{iopams}  
\usepackage{bm}
\usepackage{theorem}
\usepackage{amssymb}
\usepackage{graphicx}
\usepackage{color}

\theorembodyfont{\itshape}
\newtheorem{Theorem}{Theorem}
\theorembodyfont{\itshape}

\theorembodyfont{\itshape}

\theorembodyfont{\itshape}

\theorembodyfont{\rmfamily}
\newtheorem{Definition}[Theorem]{Definition}
\theorembodyfont{\rmfamily}

\begin{document}

\title[Modeling event cascades using networks of additive count sequences]{Modeling event cascades using networks of additive count sequences}

\author{Shinsuke Koyama$^{1,2}$ \& Yoshi Fujiwara$^3$}

\address{$^1$Department of Statistical Modeling, The Institute of Statistical Mathematics, Tokyo, Japan\\
$^2$Department of Statistical Science, Graduate University for Advanced Studies (SOKENDAI), Tokyo, Japan\\
$^3$ Graduate School of Simulation Studies, University of Hyogo, Kobe, Japan}
\ead{skoyama@ism.ac.jp}
\vspace{10pt}
\begin{indented}
\item[]July 2018
\end{indented}

\begin{abstract}
We propose a statistical model for networks of event count sequences built on a cascade structure. 
We assume that each event triggers successor events, whose counts follow additive probability distributions; the ensemble of counts is given by their superposition. 
These assumptions allow the marginal distribution of count sequences and the conditional distribution of event cascades to take analytic forms. 
We present our model framework using Poisson and negative binomial distributions as the building blocks.  
Based on this formulation, we describe a statistical method for estimating the model parameters and event cascades from the observed count sequences. 
\end{abstract}

%
%
%
%
%

\section{Introduction}

This study concerns modeling and inference of event cascades, which ensue when events cause other events to occur, thus triggering further events. 
Example events in this context include chemical reactions \cite{Kampen92}, neuronal firing \cite{Beggs03}, earthquakes \cite{Ogata88}, sending an email \cite{Fox16}, posting and sharing content on social networking services \cite{Kobayashi16,Zhao15}, and urban crime \cite{Mohler11}. 
Because event cascades are universal in a wide variety of systems, its comprehension is essential for understanding the emergence of complex phenomena \cite{Bak96,Sornette06}.

Self-exciting and mutually exciting point processes (i.e., Hawkes processes) are widely used for modeling and analyzing event sequences \cite{Hawkes71a,Hawkes71b}. 
The rate at which events occur in these models is partitioned into two components: a background rate describing an exogenous effect (e.g., trends), 
and a mutually exciting component where events trigger an increase in the process rate. 
Hawkes processes exhibit rich dynamic behavior in terms of event cascades due to the latter component \cite{Onaga14,Onaga16}. 
Social data mining has received much attention\cite{Kaya17,Zafarani14}, where modeling and inference of social networks built upon Hawkes processes form active research areas \cite{Linderman14,Rizoiu2017,Xu17,Zhou13}. 

\begin{figure}[tb]
\begin{center}
\includegraphics{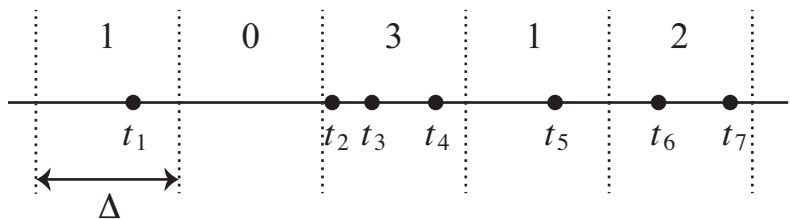}
\caption{``Coarse-graining" of event sequences. 
Event times ($t_1,t_2,\ldots$) in continuous time are aggregated within consecutive periods of length $\Delta$, resulting in a sequence of event counts ($1,0,3,1,2,\ldots$) in discrete time. 
}
\label{fig:coarse_graining}
\end{center}
\end{figure}

Whereas Hawkes processes describe a series of events in continuous time, real data are often aggregated within consecutive periods (e.g., day or week) within which the timing of each event is lost, 
resulting in sequences of event count data in discrete time (Figure~\ref{fig:coarse_graining}). 
Hawkes processes could be applied to analyze such data by neglecting the precise timing of events within each period (e.g., \cite{Lweis12}), 
but it is more desirable to use a statistical model that directly accounts for count data. 

In this paper, we propose a statistical model for count sequence networks that possesses a cascade structure. 
The key assumptions made in this model are that 
(i) the event counts triggered by preceding events follow additive probability distributions, 
and 
(ii) the ensemble of observed events is given by their superposition. 
These assumptions allow the marginal distribution of count sequences and the conditional probability distribution of the event cascades, given the count sequences, to take analytic forms.  
We illustrate our modeling framework using Poisson and negative binomial distributions, which cover a broad range of variability in event counts. 
Based on the proposed model, we develop a statistical method to estimate the event cascades and model parameters. 
The proposed method is then applied to simulated event data. 

\section{Statistical model}
\subsection{Model construction}

We consider a multivariate time series with $K$ components, 
$N_t := (n_{1t}, \ldots,n_{Kt})$ for $t \in \{t_0,t_0+1,\ldots\}$ ($t_0\in\mathbb{Z}$, being the initial time),
where $n_{it}\in\mathbb{N}_0$ represents the count of events at $(i,t)$. Here, $(i,t)$ stands for the $i$th component at time $t$.
We suppose that the $n_{it}$ events at $(i,t)$ are partitioned into two groups: events occurring because of the background rate and those triggered by preceding events. 
We make the following assumptions for these two groups of events. 
\begin{itemize}
\item[(I)] 
Let $y^b_{it}$ be the event count occurring due to the background rate, and let $P(y^b_{it})$ be its distribution. 
The expected value of $y^b_{it}$ is given by $\mathrm{E}(y^b_{it}) = \mu_{it}$.
\item[(II)] 
Let $y^c_{itjs}$ denote the count of events triggered by the preceding event at $(j,s)$ $(s<t)$, and let $P(y^c_{itjs}|n_{js})$ be its distribution, conditioned on $n_{js}$.  
The expectation of $y^c_{itjs}$ is given by
\begin{equation}
\mathrm{E}(y^c_{itjs}|n_{js}) = 
\psi_{itjs} = a_{ij}n_{js}h(t-s),
\label{eq:rate_cascade}
\end{equation}
where $a_{ij} (\ge0)$ is the strength of the influence from the $j$th to $i$th component, and $h(t)$ is the kernel function satisfying $h(t)=0$ for $t<0$ (i.e., causality) and $\sum_{t=1}^{\infty}h(t)=1$.
\end{itemize} 
Because the total number of events at $(i,t)$ is given by $n_{it}$, the following equality holds: 
\begin{equation}
n_{it} = y^b_{it} + \sum_{j=1}^K \sum_{s=t_0}^{t-1} y^c_{itjs}. 
\label{eq:n-y}
\end{equation}
Throughout this paper, we use the following notation:
\begin{eqnarray}
Y_{it} &=& \{y^b_{it}, y^c_{itjs} \ | \ j=1,\ldots,K,\ s=t_0,\ldots,t-1 \}, \nonumber\\
Y_t &=& \{Y_{it} \ | \ i=1,\ldots,K \}, \nonumber\\
Y_{s:u} &=& \{ Y_t \ | \ t=s,\ldots,u \}, \quad (s<u). \nonumber
\end{eqnarray}
We also make the following assumption:
\begin{itemize}
\item[(III)]
Given the preceding events, $Y_{t_0:t-1}$, the event counts at time $t$,  $y^b_{it}$, and $y^c_{itjs}$ are statistically independent for $i$, $j$, and $s$. 
\end{itemize}

From assumptions (I)--(III), the probability distribution of $Y_t$, 
conditioned on the preceding events $Y_{t_0:t-1}$ is 
\begin{eqnarray}
P(Y_t | Y_{t_0:t-1}) &=& 
\prod_{i=1}^K P(y^b_{it}) \prod_{j=1}^K \prod_{s=t_0}^{t-1} P(y^c_{itjs}|n_{js}). 
\label{eq:transition_prob}
\end{eqnarray}
Given an initial probability distribution of $Y_t$ at time $t=t_0$,
\begin{equation}
P(Y_{t_0}) = \prod_{i=1}^K P(y^b_{it_0}),
\label{eq:initial}
\end{equation}
the joint probability distribution of the complete data $Y_{t_0:T}$ is obtained as 
\begin{eqnarray}
P(Y_{t_0:T}) &=&
P(Y_{t_0})  \prod_{t=t_0+1}^T P(Y_t | Y_{t_0:t-1} ) \nonumber\\
&=&
\prod_{i=1}^K \prod_{t=t_0}^T P(y^b_{it}) 
\prod_{j=1}^K \prod_{s=t_0}^{t-1} P(y^c_{itjs}|n_{js}). 
\label{eq:prob_y}
\end{eqnarray}

When only the event counts at each $(i,t)$ are observed, we require the probability distribution of $N_{t_0:T}$, which is obtained from Eq.~(\ref{eq:prob_y}) as follows. 
Let $\mathcal{Y}_{it}$ be a set of $Y_{it}$ satisfying Eq.~(\ref{eq:n-y}),
\begin{equation}
\mathcal{Y}_{it} = \left\{ Y_{it}\ \Big| \ n_{it} = y^b_{it} + \sum_{j=1}^K \sum_{s=t_0}^{t-1} y^c_{itjs}
\right\},
\label{eq:setY_it}
\end{equation}
and define
\begin{eqnarray}
\mathcal{Y}_t = \{ \mathcal{Y}_{it}\ |\ i=1,\ldots,K \}, \quad
\mathcal{Y} = \{ \mathcal{Y}_t\ |\  t=t_0,\ldots,T \}. 
\label{eq:setY}
\end{eqnarray}
Then, the probability distribution of $N_{t_0:T}$ is obtained by marginalizing Eq.~(\ref{eq:prob_y}) over $\mathcal{Y}$, 
\begin{equation}
P(N_{t_0:T}) = \sum_{\mathcal{Y}} P(Y_{t_0:T}).
\label{eq:prob_n_def}
\end{equation}
The summation in the right hand side of Eq.~(\ref{eq:prob_n_def}) is generally difficult to calculate. 
However, it can be calculated analytically if the probability distributions $P(y^b_{it})$ and $P(y^c_{itjs}|n_{js})$ are {\it additive}.
\begin{Definition} \label{def:additive}
A family of probability distributions $f(y;\lambda)$ is called {\it additive} if the distribution of the sample sum $y = y_1+ \cdots + y_n$ for a random sample of size $n$ from $f(y_i;\lambda_i)$ belongs to the family itself with the parameter 
$\lambda=\lambda_1+ \cdots + \lambda_n$.
\end{Definition}

\begin{Theorem}\label{thm:main}
If $y^b_{it}$ and $y^c_{itjs}$ follow additive probability distributions, $P(y^b_{it})=f(y^b_{it};\mu_{it})$ and $P(y^c_{itjs}|n_{js})=f(y^c_{itjs};\psi_{itjs})$, respectively, and $P(N_{t_0:T})$ becomes 
\begin{equation}
P(N_{t_0:T}) = \prod_{i=1}^K \prod_{t=t_0}^T f(n_{it};\lambda_{it}), 
\label{eq:prob_n}
\end{equation}
where
\begin{eqnarray}
\lambda_{it} = 
\mu_{it} + \sum_{j=1}^K\sum_{s=t_0}^{t-1} a_{ij}n_{js}h(t-s).
\label{eq:lambda}
\end{eqnarray}
\end{Theorem}
A proof of this theorem is given in \ref{appendix:proof_main_theorem}.

Given $N_{t_0:T}$, the conditional probability distribution of $Y_{t_0:T}$ can also be factorized as follows using the additive probability distributions:
\begin{eqnarray}
P(Y_{t_0:T}|N_{t_0:T}) &=& 
\frac{P(Y_{t_0:T})}{P(N_{t_0:T})} \nonumber\\ 
&=&
\prod_{i=1}^K \prod_{t=t_0}^T
\frac{ f(y^b_{it}; \mu_{it}) }{ f(n_{it};\lambda_{it}) }
\prod_{j=1}^K \prod_{s=t_0}^{t-1} f(y^c_{itjs}; \psi_{itjs}).
\label{eq:prob_y|n}
\end{eqnarray}

\subsection{Stability condition} \label{sec:stability}

The stability condition for our model is derived as follows. 
Assume the process started a long time before ($t_0\to -\infty$), and
let $\langle \lambda \rangle_{it} = \mathrm{E}[\lambda_{it}]$ denote the expectation of the rate. 
The expectation of Eq.~(\ref{eq:lambda}) leads to
\begin{eqnarray}
\langle\lambda\rangle_{it} &=& 
\mathrm{E}\left[
\mu_{it} + \sum_{j=1}^K\sum_{s=-\infty}^{t-1} a_{ij}n_{js}h(t-s)
\right] \nonumber\\
&=& 
\mu_{it} + \sum_{j=1}^K a_{ij} \sum_{s=-\infty}^{t-1} \langle\lambda\rangle_{js} h(t-s).
\label{eq:stationarity}
\end{eqnarray}
The Z-transform of Eq.~(\ref{eq:stationarity}) is 
\begin{equation}
\Lambda_i(z) = M_i(z) + \sum_{j=1}^K a_{ij} \Lambda_j(z)H(z). 
\label{eq:z-transform}
\end{equation}
Eq.~(\ref{eq:z-transform}) can be rewritten in vector form as 
$\bm{\Lambda}(z) = \bm{M}(z) + A\bm{\Lambda}(z)H(z)$, where $A=(a_{ij})$ is the influence matrix, from which the Z-transform of the expected rate is obtained:
\begin{equation}
\bm{\Lambda}(z) = [I-AH(z)]^{-1}\bm{M}(z).
\label{eq:z-transform_rate}
\end{equation}
Thus, the spectral radius of $A$, defined by the maximum of the absolute value of the eigenvalues of $A$, must be smaller than unity in order for the expected rate to be finite. 

Indeed, under this condition, Eq.~(\ref{eq:z-transform_rate}) is expressed as 
\begin{equation}
\bm{\Lambda}(z) = \sum_{k=0}^{\infty}A^kH^k(z)
\bm{M}(z).
\end{equation}
The expected rate is obtained using the inverse of Z-transform,
\begin{equation}
\langle \bm{\lambda} \rangle_t = 
\sum_{k=0}^{\infty}A^k \underbrace{h(t)*\cdots *h(t)}_k *\bm{\mu}_t,
\end{equation}
where `$*$' represents convolution.

\subsection{Additive probability distributions}

We provide two additive probability distributions that can be used in our modeling framework. 

\subsubsection{Poisson distribution}

It is well known that the Poisson distribution has an additive property. 
The probability distribution function of a Poisson distribution is
\begin{equation}
f(y;\lambda) = \frac{\lambda^y e^{-\lambda}}{y!}, \quad \lambda>0.
\label{eq:Poisson}
\end{equation}
The mean and variance are given by $\mathrm{E}(y)=\mathrm{Var}(y) = \lambda$; the equality of the mean and variance is an important property of the Poisson distribution. 
The conditional probability distribution (\ref{eq:prob_y|n}) is derived as a product of multinomial distributions, i.e.,
\begin{eqnarray}
P(Y_{t_0:T}|N_{t_0:T})
=
\prod_{i=1}^K\prod_{t=t_0}^T
\frac{n_{it}!}{y^b_{it}!\prod_{j=1}^K\prod_{s=t_0}^{t-1}y^c_{itjs}!} 
\tilde{\mu}_{it}^{y^b_{it}} 
\prod_{j=1}^K\prod_{s=t_0}^{t-1} \tilde{\psi}_{itjs}^{y^c_{itjs}},
\label{eq:multinomial}
\end{eqnarray}
where $\tilde{\mu}_{it} = \mu_{it}/\lambda_{it}$ and $\tilde{\psi}_{itjs} = \psi_{itjs}/\lambda_{it}$. 
The conditional mean and variance of each element of $Y_{t_0:T}$ are, respectively, given by
\begin{equation}
\mathrm{E}(y^b_{it}|N_{t_0:T}) = n_{it}\tilde{\mu}_{it}, \quad 
\mathrm{Var}(y^b_{it}|N_{t_0:T}) = n_{it}\tilde{\mu}_{it}(1-\tilde{\mu}_{it}),
\label{eq:mean_var_multinomial_1}
\end{equation}
and 
\begin{equation}
\mathrm{E}(y^c_{itjs}|N_{t_0:T}) = n_{it}\tilde{\psi}_{itjs}, \quad
\mathrm{Var}(y^c_{itjs}|N_{t_0:T}) = n_{it}\tilde{\psi}_{itjs}(1-\tilde{\psi}_{itjs}).
\label{eq:mean_var_multinomial_2}
\end{equation}

Inherent in the Poisson distribution is the requirement that events are independent of one another. 
Thus, the Poisson distribution would be adequate for modeling coarse-grained data when successive events are independent of each other.

\subsubsection{Negative binomial distribution}

We consider the negative binomial (NB) distribution in the following form: 
\begin{equation}
f(y;\lambda,\phi) = 
\frac{\Gamma(y+\frac{\lambda}{\phi})}{\Gamma(y+1)\Gamma(\frac{\lambda}{\phi})}
\left(\frac{\phi}{1+\phi}\right)^y \left(\frac{1}{1+\phi} \right)^{\frac{\lambda}{\phi}} ,
\quad \lambda>0,\ \phi>0, 
\label{eq:NB}
\end{equation}
where $\Gamma( \cdot )$ is the Gamma function.
The properties of the NB distribution are summarized in \ref{appendix:NB}.
The NB distribution is additive 
with mean and variance given by $\mathrm{E}(y) = \lambda$ and $\mathrm{Var}(y) = (1+\phi)\lambda$, respectively. 
Note that the variance is greater than the mean, 
and the extra variability is controlled by $\phi$. 
The NB distribution converges to a Poisson distribution as $\phi \to 0$.
The conditional probability distribution (\ref{eq:prob_y|n}) using the NB distribution is derived as a product of Dirichlet-multinomial (DM) distributions as follows: 
\begin{eqnarray}
P(Y_{t_0:T}|N_{t_0:T}) &=& 
\prod_{i=1}^K\prod_{t=t_0}^T
\frac{\Gamma(n_{it}+1)\Gamma(\frac{\lambda_{it}}{\phi})}{\Gamma(n_{it}+\frac{\lambda_{it}}{\phi})} 
\frac{\Gamma(y^b_{it}+\frac{\mu_{it}}{\phi})}{\Gamma(y^b_{it}+1)\Gamma(\frac{\mu_{it}}{\phi})} \nonumber\\
& & { } \times
\prod_{j=1}^K\prod_{s=t_0}^{t-1}
\frac{\Gamma(y^c_{itjs}+\frac{\psi_{itjs}}{\phi})}{\Gamma(y^c_{itjs}+1)\Gamma(\frac{\psi_{itjs}}{\phi})} .
\label{eq:Dirichlet_multinomial}
\end{eqnarray}
The conditional mean and variance of each element of $Y_{t_0:T}$ are, respectively, given by
\begin{equation}
\mathrm{E}(y^b_{it}|N_{t_0:T}) = n_{it}\tilde{\mu}_{it}, \quad 
\mathrm{Var}(y^b_{it}|N_{t_0:T}) = \kappa_{it} n_{it}\tilde{\mu}_{it}(1-\tilde{\mu}_{it}),
\end{equation}
and 
\begin{equation}
\mathrm{E}(y^c_{itjs}|N_{t_0:T}) = n_{it}\tilde{\psi}_{itjs}, \quad
\mathrm{Var}(y^c_{itjs}|N_{t_0:T}) = \kappa_{it} n_{it}\tilde{\psi}_{itjs}(1-\tilde{\psi}_{itjs}),
\end{equation}
where 
\begin{equation}
\kappa_{it} = \frac{\lambda_{it} + \phi n_{it}}{\lambda_{it}+\phi} \  (> 1).
\end{equation}
Compared with Eqs.~(\ref{eq:mean_var_multinomial_1}) and (\ref{eq:mean_var_multinomial_2}), we see that the variance of the DM distribution is greater than that of the multinomial distribution. 
The DM distribution (\ref{eq:Dirichlet_multinomial}) converges to the multinomial distribution (\ref{eq:multinomial}) as $\phi \to 0$.
The properties of the DM distribution are summarized in \ref{appendix:DM}.

The Poisson assumption is violated if events positively correlate with each other, resulting in over-dispersion characterized by the count variance being greater than the mean. 
Therefore, the NB distribution may be appropriate when successive events are positively correlated. 

Note that the NB distribution with a value of $\phi$ close to zero is statistically indistinguishable from the Poisson distribution.
In this sense the Poisson distribution is a variety of the NB distribution for $\phi=0$.

\subsection{Inference of event cascades}

We consider a situation in which only the event counts $N_{1:T}$ are given for the data, and $Y_{1:T}$ are treated as latent variables. 
Thus, we wish to estimate $Y_{1:T}$ from $N_{1:T}$. 
For simplicity, we assume that the background rate $\mu_{it}=\mu_i$ is constant in time. 
We express the probability distributions as $P(N_{1:T}; \Theta,\phi)$ and $P(Y_{1:T}|N_{1:T} ; \Theta,\phi)$, 
where $\Theta := \{\bm{\mu},A\}$ is the set of parameters in $\lambda_{it}$ (Eq.~(\ref{eq:lambda})). 
The estimation method consists of two steps: 
(i) estimate the parameters $\Theta$ and $\phi$, 
and 
(ii) estimate the latent variables $Y_{1:T}$ using the estimated parameters $\hat{\Theta}$ and $\hat{\phi}$.

The parameters are estimated from the data based on the conventional maximum likelihood (ML) principle.
The log-likelihood function of the parameters is expressed using the additive probability distribution as follows:
\begin{eqnarray}
l(\Theta,\phi; N_{1:T}) &=& 
\log P(N_{1:T}; \Theta,\phi) \nonumber\\
&=&
\sum_{i=1}^K \sum_{t=1}^T \log f(n_{it};\lambda_{it}(\Theta), \phi), 
\label{eq:loglikelihood}
\end{eqnarray}
and its derivatives are 
\begin{eqnarray}
\frac{\partial l}{\partial \theta} &=&
\sum_{i=1}^K \sum_{t=1}^T
\left[
\frac{\partial}{\partial \lambda_{it}} \log f(n_{it};\lambda_{it},\phi)
\right]
\frac{\partial \lambda_{it}}{\partial \theta}, \quad \theta \in \Theta, 
\nonumber\\
\frac{\partial l}{\partial \phi} &=&
\sum_{i=1}^K \sum_{t=1}^T 
\frac{\partial}{\partial \phi} \log f(n_{it};\lambda_{it},\phi), \quad \phi>0,
\end{eqnarray}
where $f(n_{it};\lambda_{it},\phi)$ is given by the NB distribution (\ref{eq:NB}) for $\phi>0$ and by the Poisson distribution (\ref{eq:Poisson}) for $\phi=0$.
The optimal parameters $\hat{\Theta}$ and $\hat{\phi}$ are determined by maximizing the log-likelihood function under the constraint where $\Theta$ and $\phi$ are non-negative.
This optimization can be performed using standard numerical techniques \cite{Press92}.

With the estimated parameters, 
the latent variables are estimated based on the conditional probability distribution, 
$P(Y_{1:T}|N_{1:T}; \hat{\Theta},\hat{\phi})$. 
The conditional expectation 
$\hat{Y}_{1:T} = \mathrm{E}(Y_{1:T}|N_{1:T}; \hat{\Theta},\hat{\phi})$ 
provides an estimate with minimum squared error \cite{Berger93}.

\section{Simulation study}

\begin{figure}[tb]
\begin{center}
\includegraphics[width=14cm]{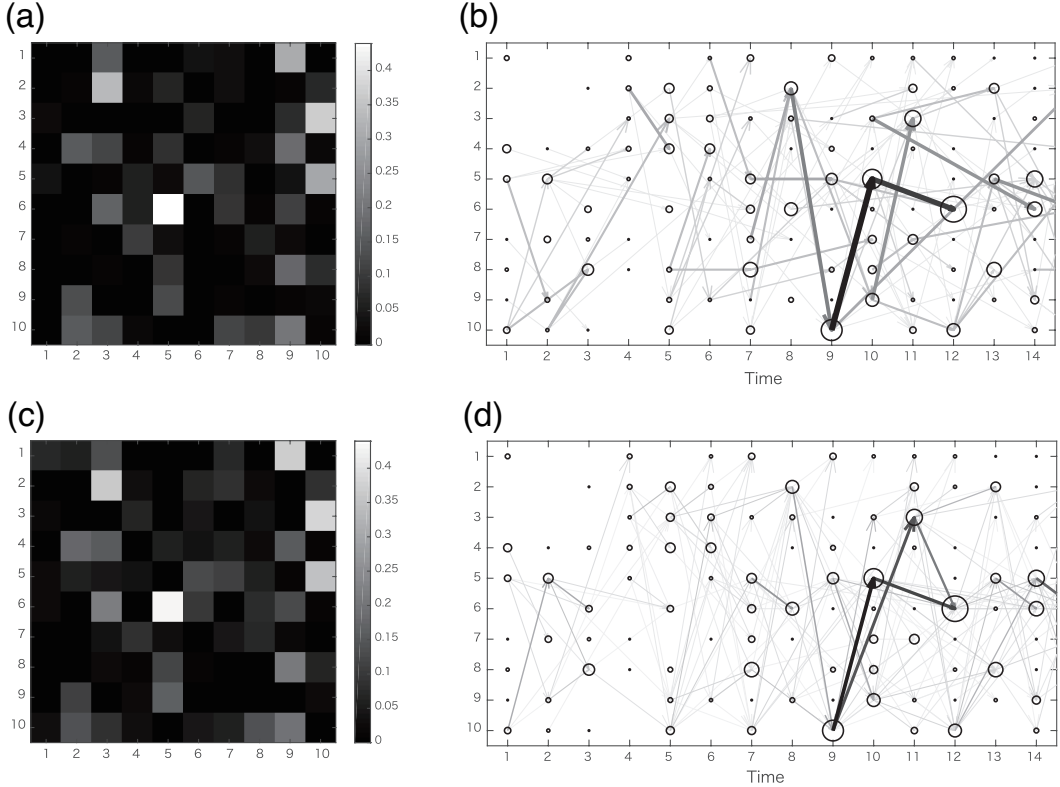}
\caption{
(a) Matrix $A$ used for simulation and (b) sample of event counts $\{n_{it}\}$ (circle) and cascades $\{y^c_{itjs}\}$ (arrow). 
The diameter of the circle and width of the arrow are proportional to the event counts. 
(c) Estimated matrix $\hat{A}$ and (d) estimated cascades $\hat{Y}_{1:14}$, which were estimated from data in $T=10^3$.
}
\label{fig:A_Y}
\end{center}
\end{figure}

We applied our method to synthetic data in order to examine the extent to which our method can extract event cascades. 
We generated data from the probability distribution (\ref{eq:prob_y}) with $K=10$ components using the NB distribution.
We used an exponential function for the kernel 
$h(t) = ce^{-t/\tau}$ ($c = 1/\sum_{t=1}^{\infty}e^{-t/\tau}$)
with time constant $\tau=2$.
The background rates were set to $\mu_i = 5$ for $i=1,\ldots,K$. 
The elements $a_{ij}$ of the matrix $A$ were generated from a gamma distribution whose mean and shape parameters were 0.05 and 0.4, respectively (Figure~\ref{fig:A_Y}a).

Simulations were performed via the following steps. 
First, the model was simulated over a time interval of $t=1$ to $T$ in order to generate samples for $Y_{1:T}$ and $N_{1:T}$ (Figure~\ref{fig:A_Y}b). 
We then estimated the parameters $\phi$, $\tau$, $\bm{\mu}$, and $A$ using the ML method, and the latent variable $Y_{1:T}$ was estimated from $N_{1:T}$ (Figure~\ref{fig:A_Y}cd).
We repeated these steps while varying the simulation interval $T$ and dispersion parameter $\phi$.

The estimation performance was quantified using the mean-square error (MSE) between the true and estimated parameters. 
To compute the MSE for the parameter $\theta \in \{\phi, \tau,\bm{\mu}, A\}$, we performed the simulation with $M=100$ repetitions for each set of parameter values. 
Denoting the estimate in the $i$th repetition by $\hat{\theta}^{(i)}$, the MSE was computed as follows:
\begin{eqnarray}
\mathrm{MSE} &:=& 
\frac{1}{M}\sum_{i=1}^M \| \theta -\hat{\theta}^{(i)} \|^2  \nonumber\\
&=&
\left\| \theta - \frac{1}{M}\sum_{j=1}^M\hat{\theta}^{(j)} \right\|^2 + 
\frac{1}{M}\sum_{i=1}^M
\left\| \hat{\theta}^{(i)} - 
 \frac{1}{M}\sum_{j=1}^M\hat{\theta}^{(j)}
 \right\|^2, 
\label{eq:MSE}
\end{eqnarray}
where $\| \cdot \|$ represents the Euclidean norm (i.e., the Frobenius norm for $\theta = A$). 
The first and second terms on the second line of Eq.~(\ref{eq:MSE}) are the bias and variance, respectively.

\begin{figure}[tb]
\begin{center}
\includegraphics[width=14cm]{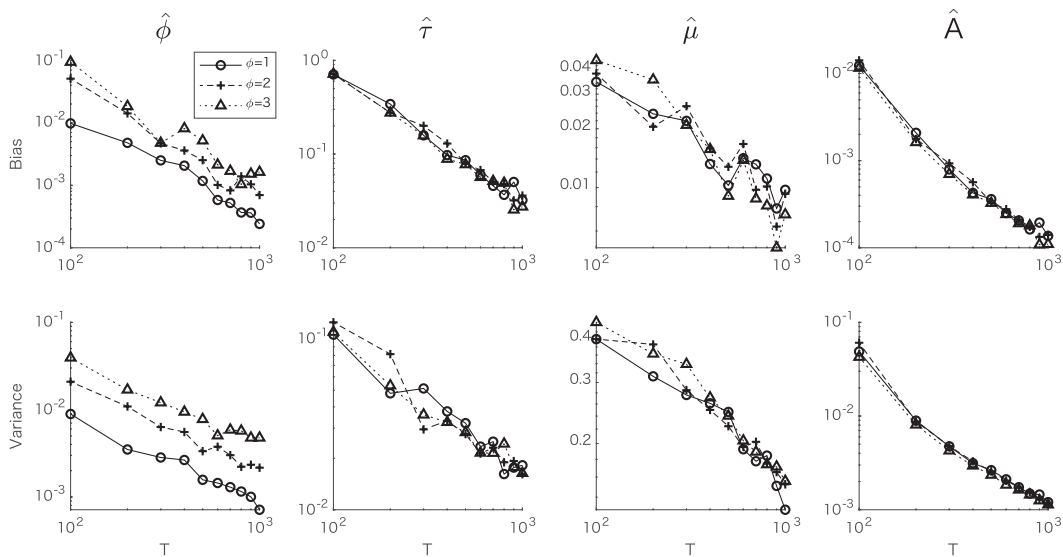}
\caption{Bias (top) and variance (bottom) in $\hat{\phi}$, $\hat{\tau}$, $\hat{\bm{\mu}}$, and $\hat{A}$ as a function of the simulation interval $T$.
Circles, crosses, and upward triangles represent those for $\phi=1$, $2$, and $3$, respectively. 
The results in this figure were computed by averaging results from 100 simulations. 
The bias and variance decrease as $T$ increases. 
}
\label{fig:error_parameters}
\end{center}
\end{figure}

The results are summarized in Figure~\ref{fig:error_parameters}. 
We see that bias and variance decrease as $T$ increases. 
The bias is an order of magnitude smaller than the variance in $\hat{\bm{\mu}}$ and $\hat{A}$; those values are comparable in $\hat{\phi}$. 
However, the bias in $\hat{\tau}$ is greater than the variance; this indicates that the estimate of the time constant is relatively less accurate. 

\begin{figure}[tb]
\begin{center}
\includegraphics[width=14cm]{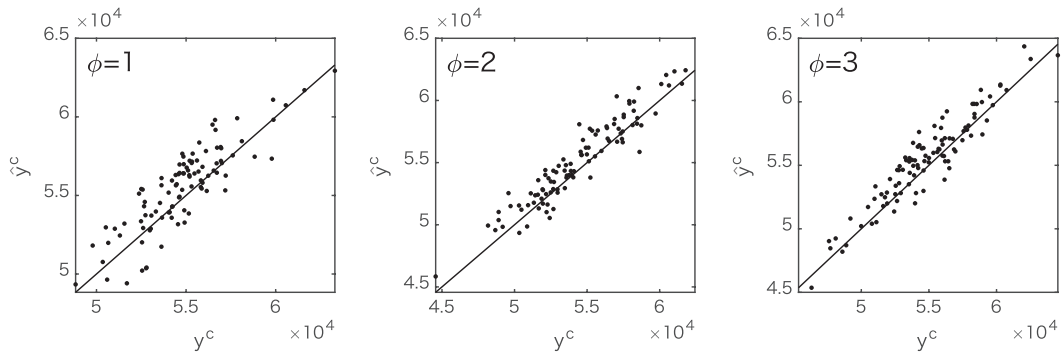}
\caption{
Scatter plot of the total number of expected triggered events $\hat{y}^c$ against the true value $y^c$ for $\phi=1$ (left), $\phi=2$ (center), and $\phi=3$ (right). 
Points approximately lie along a diagonal line, indicating that the total number of triggered events is estimated well. 
}
\label{fig:scatter_yc}
\end{center}
\end{figure}

Once we determine the optimal parameter values, we can compute the conditional expectation of the latent variables $\hat{Y}_{1:T}$, from which the detailed statistical characteristics of the event cascades can be extracted. 
The total number of triggered events is estimated as follows:
\begin{equation}
\hat{y}^c := \sum_{i=1}^K\sum_{t=1}^T\sum_{j=1}^K\sum_{s=1}^{t-1} \hat{y}^c_{itjs}.
\end{equation}
Figure~\ref{fig:scatter_yc} shows a scatter plot of $\hat{y}^c$ against the true value, from which we confirm that the number of triggered events is estimated reasonably well. 

\begin{figure}[tb]
\begin{center}
\includegraphics[width=14cm]{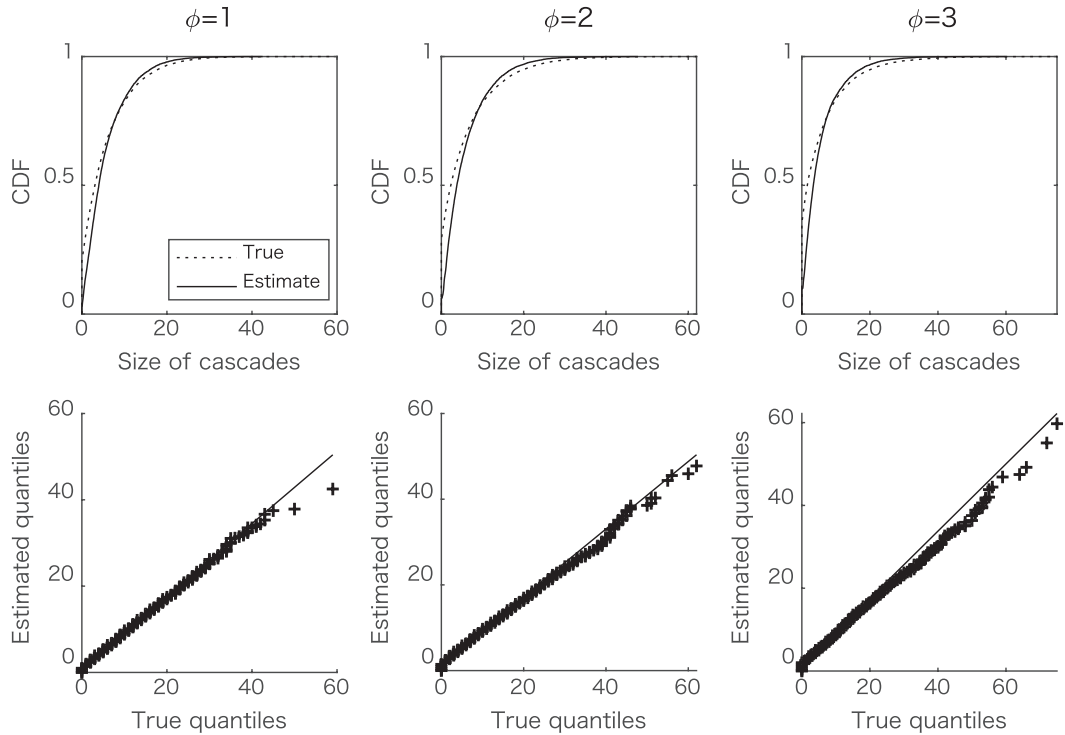}
\caption{
Top: Empirical cumulative distribution function of the size of event cascades (dotted line) and that of the estimated size of event cascades (solid line) 
for $\phi=1$ (left), $\phi=2$ (middle), and $\phi=3$ (right). 
Bottom: Corresponding Q-Q plot of the two distributions. 
Points approximately lie along a diagonal line, indicating that two distributions agree. 
}
\label{fig:csize_dist}
\end{center}
\end{figure}

We may define the ``size of event cascades" as
\begin{equation}
\hat{y}^c_{js} := 
\sum_{i=1}^K\sum_{t=s+1}^T \hat{y}^c_{itjs},
\label{eq:size_cascade}
\end{equation} 
which represents the expected number of events triggered by events at $(j,s)$. 
Figure~\ref{fig:csize_dist} (top panel) shows an empirical cumulative distribution function of the estimated size $\{\hat{y}^c_{js} :  j=1,\ldots,K,\ s=1,\ldots,T \}$ alongside that of the true size. 
These two empirical distributions can be compared visually using a quantile-quantile (Q-Q) plot, which is constructed by plotting the quantile for the estimated size against that for the true size (Figure~\ref{fig:csize_dist}, bottom panel).
We see that the points approximately lie on a line, confirming that the two distributions agree overall. 
The disagreement in the higher estimated quantiles indicates that 
the frequency of large event cascades tends to be underestimated.

\section{Discussion}

In this study, we propose a statistical model for networks of event count sequences built on a cascade structure. The key to our modeling framework is the use of additive probability distributions as the building blocks.
Their convolution property allows the marginal distribution of the count sequences and the conditional distribution of the event cascades to take analytic forms. 
We presented our method with the two additive probability distributions: the Poisson and NB distributions. 
Using these two distributions, the conditional distributions of the event cascades are found to be multinomial and Dirichlet-multinomial distributions, respectively. 

The data (i.e., measurements) we considered form count sequences in discrete time. 
Such measurements may be obtained via ``coarse-graining" of the underlying event sequences in continuous time (Figure~\ref{fig:coarse_graining}).  
Our model becomes a Hawkes process in the continuous time limit. 
Thus, the Hawkes model works as well as our model when the number of events falling in each time window is only 0 or 1. 
The latter would be preferred to the former when the time resolution is not fine enough to resolve individual events.

In general, our model may be validated against real data when the time resolution of the data series is finer than the time scale of event cascades inherent in the underlying point processes. 
When this is not the case, events occurring within the same time window may no longer be independent across different nodes of the networks, violating the model assumption (III).

We demonstrated our method with simulated data, in which all entries $a_{ij}$ of the influence matrix $A$ were independent random variables drawn from the same distribution. 
In practical situations, however, the influence matrix may be structured, 
e.g., the diagonal entries of $A$ might be larger than the off-diagonal entries, reflecting the fact that a given event could more easily trigger a later event of the same type. 
We would expect that our method works just as well for estimating the structured matrix if enough data is available based on the optimality of the maximum likelihood principle. 

Further development may be required to apply this technique to analyzing real data. 
First, for large scale networks, it is necessary to develop statistical methods to estimate the matrix $A$ from a limited amount of data because the conventional ML method may fail \cite{Hastie15}. 
Second, we assumed that the background rate is constant in time, 
but such an assumption may not be valid in a situation where nonstationary effects (e.g., seasonality and trends) are not negligible \cite{Omi17}.
Inference of a time-dependent background rate will pose a challenging problem.

\section*{Acknowledgments}

This study was supported in part by MEXT as Exploratory
Challenges on Post-K computer (Studies of Multi-level
Spatiotemporal Simulation of Socioeconomic Phenomena), and
Grant-in-Aid for Scientific Research (KAKENHI) by JSPS Grant
Number 17H02041.

\appendix
\section{Proof of Theorem~\ref{thm:main}}\label{appendix:proof_main_theorem}

Notice that $P(Y_t | Y_{t_0:t-1})$ depends only on $N_{t_0:t-1}$ through $\psi_{itjs}$. Thus, it can be expressed as $P(Y_t | Y_{t_0:t-1}) = P(Y_t | N_{t_0:t-1})$.
Using Eqs.~(\ref{eq:setY_it}) and (\ref{eq:setY}), Eq.~(\ref{eq:prob_n_def}) is calculated as follows:
\begin{eqnarray}
P(N_{t_0:T}) 
&=&
\sum_{\mathcal{Y}_{t_0}} \sum_{\mathcal{Y}_{t_0+1}} \cdots \sum_{\mathcal{Y}_T} 
P(Y_{t_0:T})
\nonumber\\
&=&
\sum_{\mathcal{Y}_{t_0}} \sum_{\mathcal{Y}_{t_0+1}} \cdots \sum_{\mathcal{Y}_T} 
P(Y_{t_0})  \prod_{t=t_0+1}^T P(Y_t | N_{t_0:t-1} )
\nonumber\\
&=&
\prod_{t=t_0}^T 
\sum_{\mathcal{Y}_t}P(Y_t | N_{t_0:t-1} ) 
\quad (\mathrm{where}\ P(Y_{t_0}|N_{t_0:t_0-1}) := P(Y_{t_0})) 
\nonumber\\
&=&
\prod_{t=t_0}^T \sum_{\mathcal{Y}_t} \prod_{i=1}^K P(y^b_{it})
\prod_{j=1}^K \prod_{s=t_0}^{t-1} P(y^c_{itjs}|n_{js}) \nonumber\\
&=&
\prod_{t=t_0}^T \prod_{i=1}^K \sum_{\mathcal{Y}_{it}}P(y^b_{it})
\prod_{j=1}^K \prod_{s=t_0}^{t-1} P(y^c_{itjs}|n_{js}).
\label{eq:prob_n_proof}
\end{eqnarray}
Using additive probability distributions and their convolution property leads to 
\begin{eqnarray}
\sum_{\mathcal{Y}_{it}}P(y^b_{it})
\prod_{j=1}^K \prod_{s=t_0}^{t-1} P(y^c_{itjs}|n_{js}) 
&=&
\sum_{\mathcal{Y}_{it}}f(y^b_{it};\mu_{it})
\prod_{j=1}^K \prod_{s=t_0}^{t-1} f(y^c_{itjs};\psi_{itjs}) \nonumber\\
&=&
f(n_{it}|\lambda_{it}),
\label{eq:f_proof}
\end{eqnarray}
where $n_{it}$ and $\lambda_{it}$ are given by Eqs.~(\ref{eq:n-y}) and (\ref{eq:lambda}), respectively.
Substituting Eq.~(\ref{eq:f_proof}) into Eq.~(\ref{eq:prob_n_proof}) yields Eq.~(\ref{eq:prob_n}).

\section{Negative binomial distribution}
\label{appendix:NB}
Here, we summarize the properties of the NB distribution used in this paper. 
See \cite{Hilbe11,Jorgensen97} for a comprehensive review. 
The probability distribution function of an NB distribution is usually expressed in the following form:
\begin{equation}
f(y;r,p) = \frac{\Gamma(y+r)}{\Gamma(y+1)\Gamma(r)}p^y(1-p)^r,
\quad
r>0,\ 0 < p < 1, 
\label{eq:NB_original}
\end{equation}
which is conventionally interpreted as the probability of the number of successes before $r$ failures occur in a series of independent Bernoulli trials with success probability $p$. 
Note that $r$ is taken as a real number greater than 0, despite this interpretation.
The NB distribution is also derived from a Poisson-gamma mixture distribution. 
The cumulant generating function (CGF) of Eq.~(\ref{eq:NB_original}) is given by 
\begin{eqnarray}
K(s) &:=& \log\mathrm{E}(e^{sy}) \nonumber\\
&=&
r\log\frac{1-p}{1-pe^s}, 
\label{eq:CGF_NB}
\end{eqnarray}
from which the mean and variance are 
$\mathrm{E}(y)=rp/(1-p)$ and $\mathrm{Var}(y)=rp/(1-p)^2$, respectively.
By changing the parameters from $(r,p)$ to $(\lambda,\phi)$ with
\begin{equation}
\lambda = \frac{rp}{1-p}, \quad 
\phi = \frac{p}{1-p}, 
\end{equation}
we obtain Eq.~(\ref{eq:NB}). 
Accordingly, the CGF is expressed as 
\begin{equation}
K(s) = -\frac{\lambda}{\phi} \log [1 - (e^s-1)\phi]. 
\label{eq:CGF_NB_2}
\end{equation}
The additivity of the NB distribution is easily confirmed using Eq.~(\ref{eq:CGF_NB_2}) as follows: 
Suppose that $y_1,\ldots,y_n$ are independent and identically distributed with $f(y_i; \lambda_i,\phi)$. The resulting CGF of $y = y_1+\cdots+ y_n$ is given by 
\begin{equation}
K(s) = - \frac{\sum_{i=1}^n\lambda_i}{\phi} \log [1 - (e^s-1)\phi],
\end{equation}
which is the CGF of $f(y;\sum_{i=1}^n\lambda_i,\phi)$. 

Expanding Eq.~(\ref{eq:CGF_NB_2}) up to leading order in $\phi$ yields
\begin{equation}
K(s) = \frac{\lambda}{\phi}[(e^s-1)\phi + o(\phi)].
\end{equation}
Thus, we obtain the CGF $K(s)=\lambda(e^s-1)$ of the Poisson distribution (\ref{eq:Poisson}) for $\phi \to 0$.

\section{Dirichlet-multinomial distribution}
\label{appendix:DM}

In this appendix, we summarize several properties of the Dirichlet-multinomial (DM) distribution and provide additional insight. 
The probability distribution function of the DM distribution is expressed as 
\begin{equation}
f(\{y_i\}_{i=1}^n ; y, \{\alpha_i\}_{i=1}^n ) = 
\frac{\Gamma(y+1)\Gamma(\alpha)}{\Gamma(y+\alpha)}
\prod_{i=1}^n\frac{\Gamma(y_i+\alpha_i)}{\Gamma(y_i+1)\Gamma(\alpha_i)}, 
\quad \alpha_i>0,
\label{eq:DM_original}
\end{equation}
where $y=y_1 + \cdots + y_n$ and $\alpha = \alpha_1 + \cdots + \alpha_n$.
The mean and variance of $y_i$ are given by 
\begin{equation}
\mathrm{E}(y_i) = y\frac{\alpha_i}{\alpha}, \quad 
\mathrm{Var}(y_i) =  y\frac{\alpha_i}{\alpha} \left(1- y\frac{\alpha_i}{\alpha}\right) \left( \frac{n+\alpha}{1+\alpha} \right). 
\end{equation}

The DM distribution is conventionally derived as a compound distribution of Dirichlet and multinomial distributions \cite{Ng11}. 
We provide another derivation here. 
Let $y_1,\ldots,y_n$ be independent and identically distributed with additive probability distributions $f(y_i; \lambda_i)$. 
From the additivity property, $y=y_1+\cdots+y_n$ follows $f(y; \lambda)$ with $\lambda=\lambda_1+\cdots+\lambda_n$. 
Given that $y=y_1+\cdots+y_n$, the conditional distribution of $\{y_i\}_{i=1}^n$ is
\begin{equation}
P(\{y_i\}_{i=1}^n | y) = \frac{\prod_{i=1}^nf(y_i; \lambda_i)}{f(y; \lambda)}. 
\label{eq:conditional}
\end{equation}
Using the NB distribution (\ref{eq:NB}) for $f(y;\lambda)$, Eq.~(\ref{eq:conditional}) becomes 
\begin{eqnarray}
P(\{y_i\}_{i=1}^n | y) =
\frac{\Gamma(y+1)\Gamma(\frac{\lambda}{\phi})}{\Gamma(y+\frac{\lambda}{\phi})}
\prod_{i=1}^n\frac{\Gamma(y_i+\frac{\lambda_i}{\phi})}{\Gamma(y_i+1)\Gamma(\frac{\lambda_i}{\phi})},
\label{eq:DM_lambda}
\end{eqnarray}
which is the DM distribution (\ref{eq:DM_original}) with $\alpha_i=\lambda_i/\phi$. Therefore, the DM distribution is the conditional distribution derived from the NB distribution. 

Note that if we use the Poisson distribution (\ref{eq:Poisson}) for $f(y;\lambda)$, the conditional distribution (\ref{eq:conditional}) becomes a multinomial distribution:
\begin{equation}
P(\{y_i\}_{i=1}^n | y) = 
\frac{y!}{\prod_{i=1}^n y_i!}\prod_{i=1}^n \left(\frac{\lambda_i}{\lambda}\right)^{y_i}.
\label{eq:multinomial_original}
\end{equation}
Thus, it follows from the convergence of the NB distribution to the Poisson distribution that the DM distribution (\ref{eq:DM_lambda}) converges to the multinomial distribution (\ref{eq:multinomial_original}) for $\phi\to0$.
Table~\ref{tbl:relation} summarizes the relationship between the four distributions. 

\begin{table}[t]
\begin{center}
  \begin{tabular}{ ccc}
    \hline\hline
    Additive distribution &  & Conditional distribution \\
    $f(y;\lambda)$    &   &  $P(\{y_i\}_{i=1}^n | y)$ \\  \hline
    Poisson (\ref{eq:Poisson}) &  $\Rightarrow$ & Multinomial  (\ref{eq:multinomial_original}) \\
    $\Uparrow$ ($\phi\to0$) &   & $\Uparrow$ ($\phi\to0$) \\
    Negative binomial (\ref{eq:NB}) & $\Rightarrow$ & Dirichlet-multinomial (\ref{eq:DM_lambda}) \\
    \hline\hline
  \end{tabular}
\end{center}
\caption{Relationship between the four distributions. }
\label{tbl:relation} 
\end{table}

\section*{References}
\bibliographystyle{plain}
\bibliography{mybib}

\end{document}